\documentclass[aps,prb,reprint,twocolumn,superscriptaddress]{revtex4-2}
\usepackage{amsmath}
\usepackage{graphicx}
\usepackage{amssymb}
\usepackage{hyperref}
\usepackage{mathtools}
\hypersetup{
    colorlinks=true,
    linkcolor=blue,
    citecolor=blue,
    urlcolor=blue
}

\begin{document}

\title{Emergent Distance and Metricity of Mutual Information in 1D Quantum Chains}
\author{Beau Leighton-Trudel}
\affiliation{Independent Researcher, Saint Petersburg, Florida, USA}
\date{November 3, 2025}

\begin{abstract}
We develop and formalize a phase diagnostic based on the information--distance 
$d_E = K_0 / \sqrt{I}$ (mutual information $I$), previously introduced for 
1D quantum chains. Calibrating $d_E$ by the Euclidean benchmark 
$I(r) \propto r^{-2} \mapsto d_E(r) \propto r$ renders the triangle-inequality 
test parameter-free and scale-invariant. Under the stated 1D site-averaged, 
monotone scaling conditions, we establish a criterion linking decay of 
$I(r)$ to metric behavior of $d_E(r)$: power laws $I(r) \sim r^{-X}$ with 
$0 < X \le 2$ yield subadditivity (metric scaling), whereas exponential 
clustering leads to superadditivity. As an analytic check complementing our 
earlier numerical study, we verify these predictions in the 1D 
transverse-field Ising chain using an exact Jordan--Wigner/BdG solution: 
at criticality, $I(r)$ follows a power law close to the $X = 2$ benchmark 
and the equal-legs triangle defect 
$\Delta(r,r) = d_E(2r) - 2d_E(r)$ is asymptotically non-positive; in gapped 
regimes $I(r)$ decays exponentially and $\Delta(r,r) \gg 0$. The result is a 
practical, falsifiable large-scale diagnostic that uses only site-averaged 
two-site MI and avoids geodesic reconstruction. We discuss scope (basis 
choice, finite-size windows) and provide code/data for replication.
\end{abstract}

\maketitle

\section{Introduction}

In one dimension, the mutual information between distant intervals decays algebraically at conformal critical points~\cite{CalabreseCardyTonni2009,CalabreseCardyTonni2011,Cardy2013}. In contrast, a nonzero spectral gap enforces exponential decay of two-point correlations~\cite{Hastings2006,NachtergaeleSims2006}. For the Jordan--Wigner (fermionic) two-mode mutual information used here this yields exponential decay of $I(r)$ (derivation in SM~G and SM~C.3; see also Ref.~\cite{Wolf2008}). Prior approaches reconstructed geometry by mapping normalized mutual information (MI) to edge lengths via a monotone decreasing function and then calculating the shortest-path metric~\cite{CCM2017}. Here we ask instead: \emph{(i)} Does a \emph{calibrated} information–distance exhibit metric scaling at large separations? and \emph{(ii)} does this metricity depend on the physical phase?

We introduce a global, falsifiable diagnostic based on the calibrated map
\[
d_E = K_0/\sqrt{I}
\]
(with MI in nats). We fix the map \emph{uniquely} by imposing a Euclidean benchmark: $I(r)\propto r^{-2}$ must map to $d_E(r)\propto r$. This removes free parameters, determines the constant $K_0$ up to units, and renders the triangle‑inequality test scale‑invariant; changing MI units rescales $K_0$ but leaves metricity unchanged (proof and uniqueness in SM~A–C). 

We prove a sharp criterion for the site‑averaged scaling function on the 1D line: if $I(r)\sim r^{-X}$, then $d_E(r)$ is subadditive (metric) iff $0<X\le 2$. Conversely, exponential decay asymptotically violates this condition. At the boundary $X=2$ the pure power law saturates the triangle defect and standard corrections to scaling set its sign (SM~C.4). We test this in the 1D transverse‑field Ising chain (TFIM, OBC) using an exact Jordan–Wigner and Bogoliubov–de Gennes treatment of the site‑averaged two‑mode MI. Throughout, “mutual information” means the fermionic (Jordan–Wigner) two‑mode MI; the theorem and calibration are basis‑agnostic in logic, but all phase‑classification statements here are in this fixed basis, for which $I(r)\sim r^{-2}$ at criticality. The diagnostic is the triangle defect $\Delta(r,r)\equiv d_E(2r)-2d_E(r)$ \ (see Fig.~1(b)). At criticality, we find power‑law MI with $X\le 2$ and $\Delta(r,r)\le 0$ asymptotically; away from criticality, MI decays exponentially and $\Delta(r,r)\gg 0$. Our claims are strictly global: we test the metricity of the \emph{site‑averaged} scaling function $d_E(r)$ and make no local line‑element assertions.

\section{Global metricity diagnostic: Definition, calibration, and criterion}

We formalize an \emph{extended} metric ansatz from the von Neumann mutual information (in nats) between sites $i$ and $j$, $I(i{:}j)$:
\begin{equation}
\label{eq:def_dE_pair}
d_E(i,j)\equiv
\begin{cases}
0, & i=j,\\[3pt]
+\infty, & I(i{:}j)=0 \ \text{and}\ i\neq j,\\[3pt]
K_0/\sqrt{I(i{:}j)}, & I(i{:}j)>0 \ \text{and}\ i\neq j.
\end{cases}
\end{equation}

\noindent\textit{Metric axioms (extended).}—Equation~\eqref{eq:def_dE_pair} defines a nonnegative, symmetric distance because $I(i{:}j)\ge 0$ and $I(i{:}j)=I(j{:}i)$. Identity of indiscernibles holds: $d_E(i,i)=0$. For $i\neq j$ either $I(i{:}j)>0$ giving $d_E(i,j)>0$ or $I(i{:}j)=0$ giving $d_E(i,j)=+\infty$. We adopt the standard extended‑metric conventions (e.g., $\infty+a=\infty$ and $b\le\infty$ for $a,b\ge0$) and test the triangle inequality only on triples with all three edges finite (see below).

In the states studied here the bulk‑averaged two‑site MI satisfies $I(r)>0$ for every finite separation $r$; in numerics we apply a strict MI floor and exclude sub‑floor separations (Sec.~\ref{sec:methods}; SM~D). Accordingly, we assess triangle inequalities only for the \emph{site‑averaged} distance $d_E(r)$ defined below on the domain $I(r)>0$.

To diagnose large‑scale geometry we pass from pairwise values to a \emph{site‑averaged scaling function over bulk pairs} in lattice units $(a=1)$. Let $\alpha\in(0,1/2)$ be a trimming fraction (we use $\alpha=0.15$) and define $L_{\rm trim}:=\lfloor \alpha L\rfloor$. For each separation $r$, we include pairs $(i,i{+}r)$ with
\[
i\in\bigl[L_{\rm trim},\,L{-}L_{\rm trim}{-}r\bigr),
\]
and define
\begin{equation}
\label{eq:def_scaling}
I(r)=\frac{1}{N_r}\sum_{i=i_{\min}}^{i_{\max}-1} I(i{:}i{+}r),
\qquad
d_E(r)=\frac{K_0}{\sqrt{I(r)}},
\end{equation}
with $i_{\min}=L_{\rm trim}$, $i_{\max}=L{-}L_{\rm trim}{-}r$, and $N_r=i_{\max}-i_{\min}$. Open boundaries break translation invariance at the pairwise level; bulk site averaging restores an effective translation‑invariant distance $d_E(r)=f(r)$ on $\mathbb{Z}_{\ge 0}$.

\noindent\textbf{Lemma (1D TI $\Leftrightarrow$ subadditivity).}
If $f:[0,\infty)\to[0,\infty)$ is nondecreasing with $f(0)=0$, then
$d(i,j)=f(|i-j|)$ on $\mathbb{Z}$ is a metric if and only if $f$ is subadditive:
$f(r_1+r_2)\le f(r_1)+f(r_2)$ for all $r_1,r_2\ge0$. See SM~B for the proof. We therefore \emph{explicitly restrict} our analysis to windows where the site‑averaged $d_E(r)$ is nondecreasing and verify this condition on each fit window (audit in SM~H).

\emph{Calibration and uniqueness.}—We fix the map $\Phi:I\mapsto d_E$ by a Euclidean benchmark: the power law $I(r)=C_I(2)\,r^{-2}$ ($X=2$) must map to $d_E(r)=r$. This removes free parameters and yields
\begin{equation}
\label{eq:calibration}
d_E=\Phi(I)=K_0\,I^{-1/2},\qquad K_0:=\sqrt{C_I(2)}.
\end{equation}
No other \emph{exponent‑independent} map $\Phi$ is compatible with this calibration (proof in SM~A). Changing MI units (nats $\leftrightarrow$ bits) rescales $K_0$ but, since the triangle inequality is invariant under positive rescaling, metricity and the diagnostic are unaffected; we therefore set $K_0=1$ in plots/diagnostics (Sec.~\ref{sec:methods}).

\medskip
\noindent\textbf{Proposition (Uniqueness of the Euclidean calibration).}
Let $\Phi$ be an exponent-independent map from mutual information to distance.
If the Euclidean benchmark requires that for every $C>0$,
$I(r)=C\,r^{-2}$ is mapped to $d_E(r)=r$ up to an overall unit, then
\[
\Phi(I)=K_0\,I^{-1/2}\quad\text{for some }K_0>0.
\]
Changing MI units rescales $K_0$ but does not affect (non)metricity.

\emph{Sketch of proof.}
The condition $\Phi(Cr^{-2})\propto r$ for all $C>0$ forces
$\Phi(y)=K_0\,y^{-\gamma}$ with $r^{2\gamma}$ proportional to $r$, hence $\gamma=\tfrac12$.
Any other $\gamma$ would violate the benchmark. Full details and mild regularity
assumptions are in SM~A.
\medskip

\emph{Induced scaling.}—Whenever the site‑averaged MI exhibits a power law,
\begin{equation}
I(r)=C_I(X)\,r^{-X},
\end{equation}
the calibration induces
\begin{equation}
\label{eq:scaling_power}
d_E(r)=\sqrt{\frac{C_I(2)}{C_I(X)}}\,r^{p}\equiv A(X)\,r^{p},\qquad p:=X/2.
\end{equation}
Metricity depends only on $p$; the amplitude $A(X)>0$ is immaterial. In gapped phases, exponential clustering gives $I(r)\sim C\,r^{-\kappa}e^{-r/\xi}$, hence $d_E(r)\propto r^{\kappa/2}e^{\,r/(2\xi)}$ (SM~C.3).

\noindent\textbf{Theorem 1 (Metric window for the calibrated distance).}
Let the site-averaged mutual information obey a power law
$I(r)=C_I(X)\,r^{-X}$ and fix the Euclidean calibration $d_E=\Phi(I)=K_0 I^{-1/2}$,
which induces $d_E(r)=A(X)\,r^{p}$ with $p=X/2$ and $A(X)>0$.
On the 1D line, restricted to a window where $d_E$ is nondecreasing,
$d_E$ is subadditive (and hence satisfies the triangle inequality) if and only if $0<X\le 2$.
If $I(r)$ decays exponentially, then $d_E$ is eventually superadditive.
At $X=2$ the equal-legs triangle defect $\Delta(r,r)$ cancels at leading order, and
standard corrections $I(r)=Cr^{-2}\!\left(1+c\,r^{-\omega}+\cdots\right)$ determine the sign.

\emph{Sketch of proof.}
By the Lemma, metricity reduces to subadditivity of $f(r)=A r^{p}$,
which holds precisely for $0<p\le1$ (concavity of $t^p$ on $\mathbb{R}_{\ge0}$), i.e.\ $0<X\le2$.
Exponential clustering gives $d_E(r)\propto r^{\kappa/2}\exp[r/(2\xi)]$, which violates subadditivity at large $r$.
See SM~B–C for details.

\emph{Practical diagnostic.}—We evaluate the triangle defect
\begin{equation}
\label{eq:triangle_defect}
\begin{gathered}
\Delta(r_1,r_2):=d_E(r_1{+}r_2)-d_E(r_1)-d_E(r_2),\\
\Delta(r,r)=d_E(2r)-2d_E(r).
\end{gathered}
\end{equation}
For $0<X<2$, $\Delta(r,r)\le 0$ asymptotically; for $X=2$ the leading term cancels and the sign is set by corrections (SM~C.4). Exponentially clustered regimes yield $\Delta>0$ for all sufficiently large $r$ (SM~C.3). Only above‑floor (finite) separations enter the diagnostic (Sec.~\ref{sec:methods}; SM~D).

\begin{figure*}[t]
\centering
\includegraphics[width=0.95\textwidth]{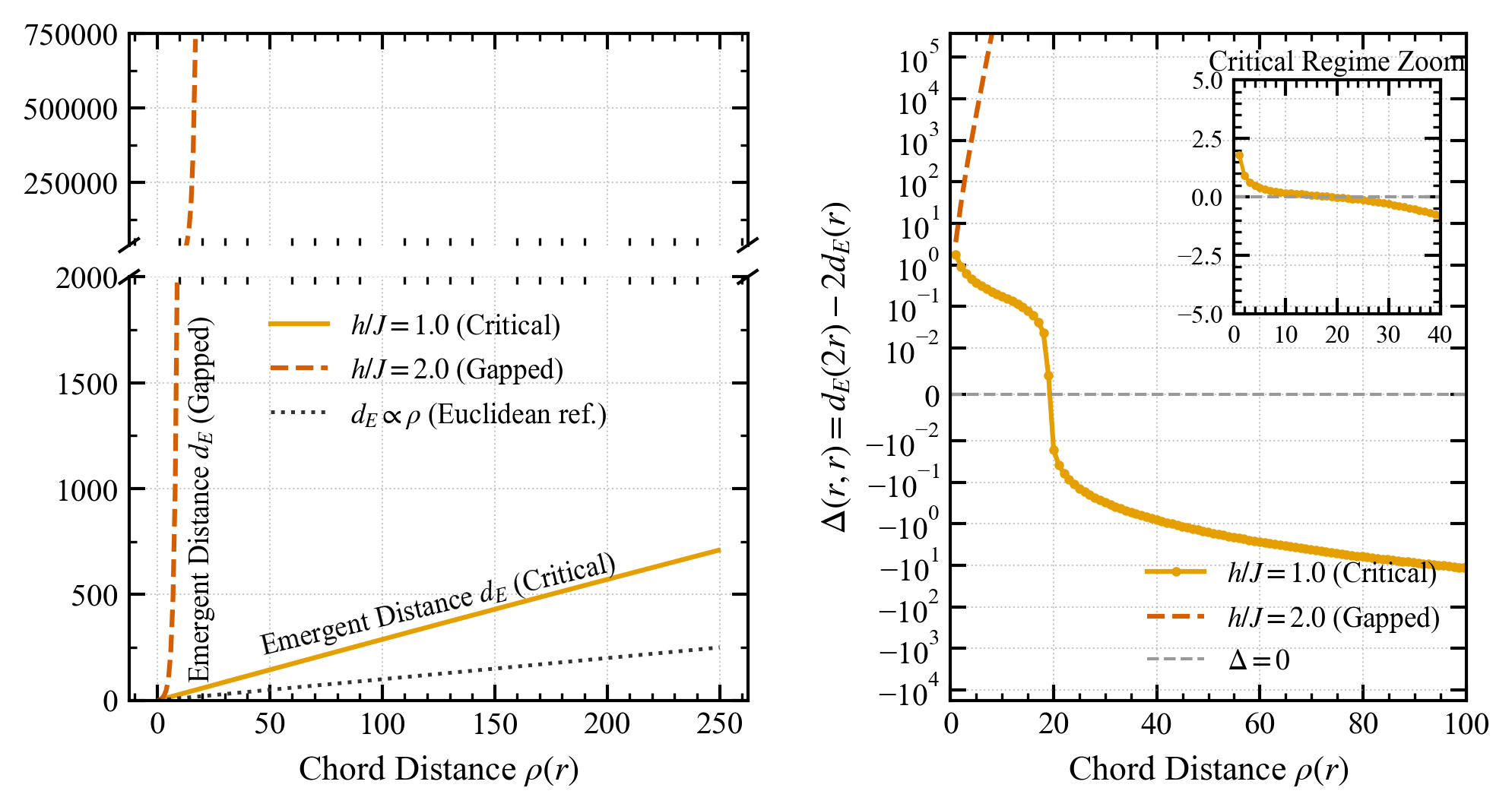}
\caption{
\textbf{Metric dichotomy in the 1D TFIM ($L=2048$).}
(a) Emergent distance $d_E=I^{-1/2}$ versus chord distance $\rho$.
At criticality ($h/J=1$) the data follow $d_E\propto\rho^{p}$ with $p=X/2=0.994(1)$ (95\% CI); the dotted line shows the Euclidean reference $p=1$.
In the gapped phase ($h/J=2$), $d_E$ grows exponentially (non‑metric).
(b) Triangle defect $\Delta(r,r)=d_E(2r)-2\,d_E(r)$ (symlog scale): the gapped phase is superadditive ($\Delta>0$), while the critical data are asymptotically subadditive ($\Delta\le0$); the small positive $\Delta$ at short distances reflects pre‑asymptotic corrections.
}
\label{fig:main_dichotomy}
\end{figure*}

\begin{figure*}[t]
    \centering
    \includegraphics[width=0.95\textwidth]{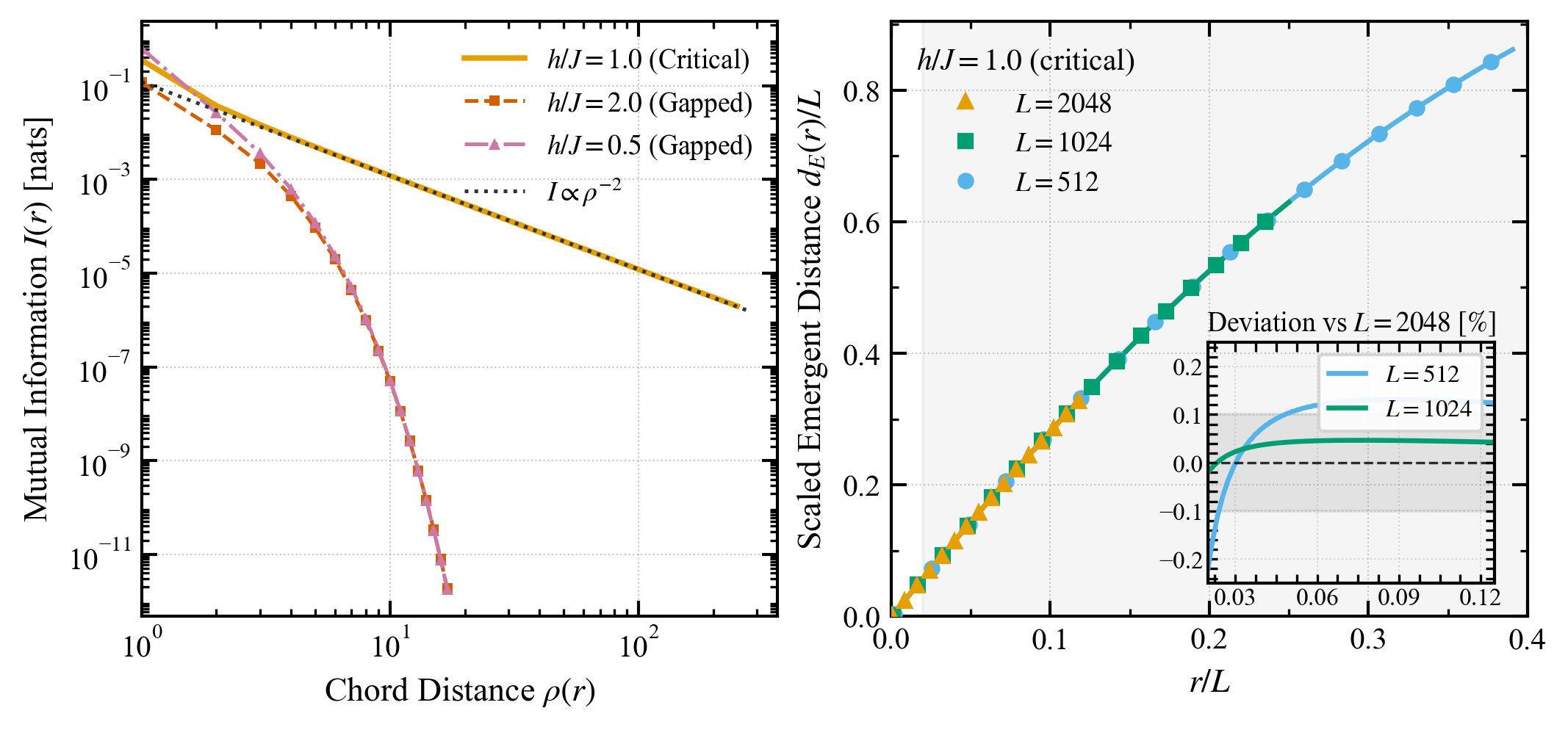}
    \caption{
\textbf{Scaling and finite-size checks.} 
\textbf{(a)} Mutual information $I(r)$ vs chord distance $\rho(r)$ for $L{=}2048$. At criticality ($h{=}J$), $I\!\propto\!\rho^{-X}$ with $X=1.988(1)$, close to the $X{=}2$ benchmark (dotted). In the gapped regimes ($h/J{=}0.5,\,2$), $I$ decays exponentially and truncates at the numerical floor ($10^{-12}$ nats).
\textbf{(b)} Critical FSS: $d_E/L$ vs $r/L$ collapses for $L\in\{512,1024,2048\}$, consistent with finite-size scaling for open chains; inset, deviation from the $L{=}2048$ curve is $\lesssim 0.5\%$ for $r/L\gtrsim 0.02$.
}
    \label{fig:scaling_and_fss}
\end{figure*}

\section{Numerical Methods} \label{sec:methods}

We study the 1D TFIM with OBC,
\begin{equation}
H=-J\sum_{i=1}^{L-1}\sigma_i^x\sigma_{i+1}^x-h\sum_{i=1}^L \sigma_i^z,
\end{equation}
on a lattice of spacing \(a=1\) (critical point \(h=J\)\,\cite{Pfeuty1970}). After a Jordan–Wigner map we solve the quadratic fermion Hamiltonian exactly by BdG\,\cite{LiebSchultzMattis1961,Pfeuty1970,BarouchMcCoy1971}, yielding the \(T{=}0\) Gaussian correlators \(C_{ij}\) and \(F_{ij}\); as a solver check, \(E_0/L\!\to\!-4/\pi\) at \(h{=}J\) (See SM~D; cf.\ \cite{Pfeuty1970,Farnell2019}).

\paragraph*{Mutual information and averaging—basis choice.}
We use the Jordan–Wigner (fermionic) two‑mode mutual information (nats) throughout, computed from the Gaussian correlation matrix \(G_A\) (SM~D, Eqs.~(S9)–(S11)). This is the observable naturally produced by our JW\(\to\)BdG solver, and it is the basis on which all phase‑classification statements in this work are made. At the TFIM critical point the Ising fermion has \(\Delta_\psi=\tfrac12\), implying \(|C_{ij}|,|F_{ij}|\sim r^{-1}\)~\cite{DiFrancesco1997}. For fermionic Gaussian states the two‑site MI is quadratic in \(C_{ij},F_{ij}\) (SM~G), hence \(I(r)\propto r^{-2}\) asymptotically~\cite{PeschelEisler2009}, consistent with our chord‑distance fits.

From \((C,F)\) we build the Nambu matrix \(G_A\) for a subsystem \(A\); its paired spectrum \((\lambda_k,1{-}\lambda_k)\) gives \(S(A)=\sum_k H_2(\lambda_k)\)\,\cite{Peschel2003,PeschelEisler2009}.
The JW two‑mode MI is \(I(i{:}j)=S(\{i\})+S(\{j\})-S(\{i,j\})\).
We form the site‑averaged \(I(r)=\langle I(i{:}i{+}r)\rangle\) over bulk pairs \(i\in[L_{\rm trim},\,L{-}L_{\rm trim}{-}r)\) with fixed trim \(\alpha=0.15\) and \(L_{\rm trim}=\lfloor\alpha L\rfloor\).
When available we record \(N_r\) and a per‑\(r\) dispersion; WLS uses \(N_r\) by default (or, if \(N_r\) is unavailable, \(1/\sigma_r^2\)); otherwise OLS.
Tiny negative MI are clipped to \(0\). With per‑pair lists we drop \(I\le 10^{-12}\) nats \emph{before} averaging—otherwise this floor acts only post‑aggregation (in \(d_E\) and fits).

\paragraph*{Diagnostics and analysis policy.}
We set \(d_E(r)=I(r)^{-1/2}\) and use the triangle defect \(\Delta(r,r)=d_E(2r)-2\,d_E(r)\).
Open boundaries break translation invariance at the pairwise level; bulk site‑averaging restores an effective translation‑invariant distance \(d_E(r)\) on \(\mathbb{Z}_{\ge0}\).
In 1D, TI\(\Leftrightarrow\)subadditivity for nondecreasing distances; we therefore restrict to fit windows where the site‑averaged series are monotone and verify this \emph{post‑run} from recorded arrays (we never enforce monotonicity by smoothing; audit in SM~H).
Floor‑masked points are excluded from \(d_E\) and all fits.
Weights enter only the exponent fits; the triangle‑defect diagnostic itself uses no regression and is independent of weighting.
Fits use the reproducible window \(r\in[12,\min(\lfloor0.35L\rfloor,256)]\) with 95\% CIs computed from the (weighted) slope standard error with a two‑sided Student‑\(t\) factor (normal fallback).

\textbf{Chord abscissa.}
At \(h{=}J\) we refit \(I(\rho)\) on the same window/weights using the PBC chord
\(\rho(r,L)=\tfrac{L}{\pi}\sin\!\big(\tfrac{\pi r}{L}\big)\)\,\cite{Calabrese2004,Calabrese2009} to reduce OBC curvature. Bulk site‑averaging strongly reduces the OBC center‑of‑mass dependence, so \(\rho\) serves as a practical abscissa to reduce curvature. Although our simulations use open boundaries, this chord refit is a curvature‑reducing \emph{change of abscissa only}: the triangle‑defect diagnostic and all metricity tests are evaluated in lattice units $r$ and are therefore independent of $\rho$ (SM~D). The SM compares the OBC variant \(\rho_{\rm OBC}(r,L)=\tfrac{2L}{\pi}\sin\!\big(\tfrac{\pi r}{2L}\big)\), related by \(\rho_{\rm OBC}=\rho/\cos\!\big(\tfrac{\pi r}{2L}\big)\); over our windows this factor is \(\approx 1.02\) at \(L=2048\) (\(\approx\)2.0\%) and \(\le 1.173\) for \(L\le 512\), and fitted exponents agree within the across‑size drift.
We quote \(X_{\rm chord}\) from the PBC refit.
The triangle defect is evaluated in lattice units, independent of this choice.

\noindent
Finite‑size scaling uses \(d_E/L\) vs \(r/L\).
For \(h\neq J\) we observe exponential decay and extract \(\xi\) from log–linear fits on the same window/weights\,\cite{Hastings2006,NachtergaeleSims2006}.

\emph{Reporting convention.} We quote exponents and correlation lengths with 2–3 significant digits; uncertainties in parentheses denote the half‑width of the 95\% confidence interval and apply to the last digit(s). Full‑precision fit values and per‑size tables are given in the SM.

\section{Results}
\label{sec:results}

\emph{Critical point ($h=J$).—} The MI exhibits a clear power law when plotted versus the chord variable, $I(\rho)\propto \rho^{-X}$, with an exponent that drifts monotonically toward the Euclidean benchmark $X=2$ as $L$ increases. The chord exponents and 95\% confidence intervals are
\[
\begin{aligned}
L{=}256:\ & X_{\mathrm{chord}}=1.963(3),\quad R^2>0.9999;\\
L{=}512:\ & X_{\mathrm{chord}}=1.968(2),\quad R^2>0.9999;\\
L{=}1024:\ & X_{\mathrm{chord}}=1.972(2),\quad R^2>0.9999;\\
L{=}2048:\ & X_{\mathrm{chord}}=1.988(1),\quad R^2>0.9999.
\end{aligned}
\]
all from a single WLS policy on $r\in[12,\,r_{\mathrm{upper}}]$. For the largest system this corresponds to $p=X/2=0.994(1)$ (95\% CI), i.e.\ within the metric window $p\le1$.

By contrast, fitting the same data versus the bare lattice distance $r$ (rather than $\rho$) underestimates the exponent due to boundary effects even at $L=2048$: $X=1.971(2)$ (95\% CI), with $R^2>0.9999$.

At criticality, $\Delta(r,r)$ becomes non-positive by $r \simeq 20$ and remains $\le 0$ for the rest of the asymptotic window; the small positive values at the shortest separations reflect pre-asymptotic corrections. For example, for $L=512$ we find $\Delta(12,12)\approx -0.21$ and it remains negative for all larger $r$ in the window. On the same windows the monotonicity audit (Sec.~\ref{sec:methods}) confirms nondecreasing $d_E(r)$ and nonincreasing $I(r)$; the worst‑case upward finite differences $\delta_{\max}^{(I)}$ and $\delta_{\max}^{(d_E)}$, their $z$‑scores, and the isotonic sup‑norm deviations are reported in Table~S2 (SM~H).

These observations are consistent with subadditivity of $d_E$ on the asymptotic window and, together with the monotonicity audit (SM~H), imply the triangle inequality on the scanned domain. As visible in Fig.~1(b, inset), $\Delta$ can be slightly positive at very short separations due to pre‑asymptotic corrections; our fits begin at $r_{\min}=12$ (see also SM~C.4 for the $X{=}2$ boundary case).

\emph{Gapped phases ($h \neq J$).—} In both paramagnetic ($h=2J$) and ferromagnetic ($h=J/2$) regimes, the site-averaged MI decays exponentially, $I(r)\propto e^{-r/\xi_{\text{MI}}}$, where the MI decay length $\xi_{\text{MI}}$ is, in general, distinct from the spin correlation length $\xi_{\text{spin}}$. In both gapped regimes we obtain short MI correlation lengths, nearly size‑independent over $L\in\{256,512,1024,2048\}$: $\xi_{\text{MI}}=0.685(2)$ at $h{=}J/2$ and $\xi_{\text{MI}}=0.689(2)$ at $h{=}2J$ (95\% CI; full per‑$L$ values are provided in the SM). As predicted for $d_E\propto e^{\,r/(2\xi_{\text{MI}})}$, the triangle defect grows rapidly and becomes strongly positive. For example, at $L=2048$, $h=2J$, $\Delta(8,8)\approx 3.53\times10^{5}$, demonstrating superadditivity (non‑metricity). An analogous growth is seen for $h=J/2$.

\emph{Finite-size scaling.—} At criticality, the scaled curves $d_E/L$ vs $r/L$ collapse across $L$, confirming that the diagnostic is governed by the asymptotic decay class of $I(r)$ rather than its nonuniversal amplitude. In the gapped phases the exponential growth precludes such a collapse under this scaling. Representative scaled arrays are included for all runs.

\emph{Global triangle test.} To rule out an equal‑legs‑only artifact, we scanned all resolvable integer pairs $(r_1,r_2)$ in the asymptotic window: at criticality ($L=2048$) $\max_{(r_1,r_2)}\Delta=\Delta(20,20)=-1.67\times10^{-2}$ ($\approx12.8\sigma$ below zero), while all resolvable gapped‑phase pairs in our window have $\Delta>0$ (see SM~I).

\emph{Summary.—} The numerical results establish a sharp, size‑robust dichotomy aligned with the theoretical criterion: at the TFIM critical point the chord–distance fit exponent approaches the metric bound ($X\!\to\!2$, $p\!=\!X/2\!\le\!1$) and $\Delta(r,r)\!\le\!0$ asymptotically; in gapped phases $I(r)$ decays exponentially, $d_E$ is superadditive, and $\Delta(r,r)\!\gg\!0$. The sign of $\Delta$ thus serves as a basis‑fixed, scale‑invariant phase diagnostic (JW two‑mode MI here), while the chord‑fit exponents quantitatively verify the Euclidean benchmark calibration of $d_E$ at criticality.

\section*{Discussion and Outlook}\label{sec:discussion}

We have introduced a \emph{global} diagnostic for quantum phases based on the geometry of correlations. The calibrated information–distance,
\[
d_E = K_0/\sqrt{I},
\]
is uniquely fixed by the Euclidean benchmark ($I\!\propto r^{-2}\mapsto d_E\!\propto r$), making the test parameter‑free and scale‑invariant. The main theorem connects the phase to geometry via a \emph{sharp} criterion: if $I(r)\sim r^{-X}$, then the site‑averaged scaling function $d_E(r)$ is a metric (subadditive) \emph{if and only if} $0<X\le 2$; exponential clustering in gapped phases yields superadditivity. Exact TFIM calculations verify this dichotomy: at criticality the exponent lies in the metric window and the triangle defect $\Delta(r,r)=d_E(2r)-2\,d_E(r)$ is asymptotically non‑positive, whereas in gapped regimes $I(r)$ decays exponentially and $\Delta(r,r)\!\gg 0$. Beyond equal legs, a full 2D scan on the asymptotic window confirms $\Delta(r_1,r_2)\le 0$ at criticality (SM~I). Thus the \emph{sign} of $\Delta$ provides a binary, falsifiable diagnostic that complements standard exponent fits.

\paragraph*{Interpretation.}
The calibration at $X=2$ is natural in $1{+}1$D (equivalently, 2D Euclidean) CFTs: in the weak‑correlation regime the two‑site MI is quadratic in the leading two‑point correlator, suggesting $X\simeq 4\Delta_\star$, where $\Delta_\star$ is the smallest scaling dimension allowed by the twist‑field OPE~\cite{Cardy2013,CalabreseCardyTonni2009,CalabreseCardyTonni2011}. For the \emph{fermionic} (JW) two‑mode MI used here, the leading operator is the Ising fermion $\psi$ with $\Delta_\psi=\tfrac12$~\cite{DiFrancesco1997}, hence $I(r)\propto r^{-2}$, consistent with Gaussian correlation‑matrix theory~\cite{PeschelEisler2009} and with our numerics. By contrast, the energy density $\varepsilon$ has $\Delta_\varepsilon=1$ and would imply $X=4>2$ and therefore non‑metric behavior. If a 1D critical system were dominated by correlations with $X>2$, our calibrated $d_E$ would be superadditive; the criterion is therefore sharp relative to the $X=2$ benchmark. 

\paragraph*{Scope and limitations.}
Our claims are deliberately \emph{global}: we probe only the bulk‑averaged scaling function $d_E(r)$ and make no assertion of a local line element or of pairwise metricity at short distances. The triangle test is applied only to finite‑edge triples (above the MI floor), and—in line with the 1D equivalence “triangle inequality $\Leftrightarrow$ subadditivity’’—only on windows where $d_E(r)$ is empirically nondecreasing (audited in the SM). The diagnostic is basis dependent: \(X\) depends on the MI definition; all positive classifications reported here (e.g., that the critical TFIM is metric) refer to the fermionic (Jordan--Wigner) two‑mode MI used throughout. The \emph{logic} of the test is basis‑agnostic: any MI with power‑law decay $X\le 2$ yields metric scaling, whereas exponential decay does not. The calibration/uniqueness proof ensures that once the Euclidean benchmark is imposed, no other exponent‑independent map $\Phi$ is compatible (SM~A). At $X=2$ the leading equal‑legs defect cancels and corrections to scaling control the sign (SM~C.4).

\paragraph*{Utility.}
Practically, the diagnostic reduces phase identification to a \emph{yes/no} question: does the calibrated $d_E(r)$ satisfy subadditivity (equivalently, is $\Delta(r,r)\le 0$ asymptotically)? It requires only site‑averaged two‑site MI and avoids fitting geodesics or extracting precise critical exponents. The analysis contract is simple to replicate, and the code/data are released to facilitate adoption. 

\paragraph*{Outlook.}
Two natural next steps, beyond the scope of this work, are: (i) applying the same analysis contract to additional exactly solvable and interacting 1D models (e.g., XY/XX, Luttinger liquids, and integrable deformations) to map the prevalence of $X\le 2$ versus exponential regimes; and (ii) exploring higher‑dimensional analogs where one would test subadditivity of appropriate \emph{radial} scaling functions rather than shortest‑path reconstructions. A second direction is basis/systematics: repeating the diagnostic with spin‑basis MI or with experimentally accessible proxies to assess robustness of the binary outcome. We anticipate the metricity test—calibrated once and used as the sign of $\Delta$—will serve as a compact, reproducible tool that complements existing entanglement‑based diagnostics across numerical studies and, where feasible, in experiments.

\bibliographystyle{apsrev4-2}
\bibliography{references}

\clearpage
\onecolumngrid
\begingroup
  \setlength{\parindent}{0pt}
  \setlength{\parskip}{0.8\baselineskip}

\section{Code and Data Availability}

All source code to generate figures, reproduce simulations, and analyze data is available at: 
\url{https://github.com/darkhorse131/emergent-geometry-metricity-TFIM}.
An archived snapshot corresponding to this version is available at 
DOI: \url{https://doi.org/10.5281/zenodo.17518193}.

\end{document}